\documentclass[%
 amsmath,amssymb,
 aps,
prl, reprint,
]{revtex4-1}

\usepackage{graphicx}
\usepackage{bm}

\begin{document}

\preprint{APS/123-QED}

\title{Transport properties of a 3D topological insulator \\ based on a strained
high mobility HgTe film}

\author{D. A. Kozlov}
 \email{dimko@isp.nsc.ru}
\author{Z. D. Kvon}
 \altaffiliation[Also at ]{Novosibirsk State University.}
 \author{E. B. Olshanetsky}
\author{N. N. Mikhailov}
\author{S. A. Dvoretsky}
 \affiliation{A. V. Rzhanov Institute of Semiconductor Physics, Novosibirsk 630090, Russia}

 \author{D. Weiss}
 \affiliation{Experimental and Applied Physics, University of Regensburg, D-93040 Regensburg, Germany}

\date{\today}

\begin{abstract}
We investigated the magnetotransport properties of strained,
80\,nm thick HgTe layers featuring a high mobility of $ \mu \sim
4\times10^5$\,cm$^2$/V$\cdot$s. By means of a top gate the
Fermi-energy is tuned from the valence band through the Dirac type
surface states into the conduction band. Magnetotransport
measurements allow to disentangle the different contributions of
conduction band electrons, holes and Dirac electrons to the
conductivity. The results are in line with previous claims
that strained  HgTe is a topological insulator with a bulk gap of
$\approx$15\,meV and gapless surface states.

\begin{description}
\item[PACS numbers]
\pacs{1} 73.25.+i, \pacs{2} 73.20.At, \pacs{3} 73.43.-f
\end{description}
\end{abstract}

\pacs{1}
\maketitle

The discovery of two (2D)- and three-dimensional (3D) topological
insulators (TI), a new material class with insulating bulk and
topologically protected, conducting surface states, has opened a
new and exciting research field in condensed matter physics
\cite{Kane05, Kane05-2, Zhang06, Zhang06-2, Kane07, Mele07,
Hasan08, Hasan09, Hasan10, Xiao11}. Although quite a number of
different, especially Bi based materials \cite{Checkelsky09,
Taskin09, Qu10, Kim12}, belong to this category, materials which
combine high charge carrier mobility and insulating bulk are still
scarce. This is mostly due to the fact that Bi-based 3D TIs are
heavily-doped alloy films with a very low mobility $\approx$
1000\,cm$^2$/V$\cdot$s and a high bulk carrier density of 10$^{17}$ -
10$^{19}$\,cm$^{-3}$. HgTe based 2D TI, on the other hand, are
characterized by very high mobilities enabling the discovery of
the quantum spin Hall effect \cite{Konig07}. A recent analysis of
the sequence of quantum Hall plateaus suggests that also strained
HgTe layers constitute a 3D TI. The strain opens a gap in the
gapless semimetal HgTe, so that the TI properties can be explored
by tuning the Fermi energy into the bulk gap and probing the
transport properties of the gapless surface states. Although the
strained HgTe film has a much higher mobility $\mu =
(3-4)\times10^4$\,cm$^2$/V$\cdot$s, the high bulk carrier density
and the absence of a top gate has complicated the detection of the
3D TI so far \cite{Brune11, Olshanetsky11}.

The strain in HgTe layers, grown by molecular beam epitaxy (MBE),
stems from  a 0.3\% lattice mismatch between HgTe and CdTe. The
corresponding critical film thickness for pseudomorphic growth is
larger than 100\,nm meaning that thinner films adopt the substrate
lattice constant. Due to this strain a small gap of $\sim 15$\,meV
opens (see below) in the bulk energy spectrum of the film. Within
the bulk gap the gapless surface states reside. The charge
neutrality point of the corresponding Dirac cone is located in the
valence band \cite{Brune11}. Thus Dirac electrons with higher
energy dominate electrical transport if the Fermi energy is
located in the bulk gap. This allows us to map the bandstructure
in the vicinity of the bulk gap by magnetotransport experiments as
is outlined below.

In this Letter we report on the transport properties of high
mobility, 80\,nm wide, strained HgTe films equipped with a gate.
The low disorder, manifested in high charge carrier mobilities,
together with the possibility to tune the Fermi energy from  the
valence via the gap into the conduction band, enables us to probe
the 2D Dirac surface states when the Fermi energy is in the energy
gap of the bulk. Since HgTe films grown on CdTe suffer from
dislocations due to the lattice mismatch, our 80\,nm thick HgTe
films were separated from the CdTe substrate by a 20\,nm thin
Cd$_{0.7}$Hg$_{0.3}$Te buffer layer. This buffer layer increases
the electron mobility by an order of magnitude (up to
$4\times10^5$\,cm$^{2}$/V$\cdot$s) and reduces the bulk impurity
concentration to values of order 10$^{16}$\,cm$^{-3}$ (see
supplemental material). We fabricated and investigated two  types
of devices: one with the upper HgTe surface uncapped and the other
one covered with a 20\,nm Cd$_{0.7}$Hg$_{0.3}$Te cap layer (see
Fig.~1(a) and supplement). Since all measured major properties
(magnetotransport traces, bulk energy gap etc.) of capped and
uncapped films were found to be similar, we will only focus on the
uncapped HgTe film below. For transport measurements the films
were patterned into Hall bars supplied with top gates.
Cross-sections of the devices are sketched in Fig.~1(a). Each Hall
bar consists of three 50\,$\mu$m wide segments of different
lengths (100, 250, and 100\,$\mu$m) with eight voltage probes.
Ohmic contacts to the active layer were formed by alloying indium.
For gating two types of dielectric layers were used, giving
similar results: 100\,nm SiO$_2$ and 200\,nm of Si$_3$N$_4$ grown
by plasma chemical vapor deposition of SiH$_4$ + N$_2$O at
100$^{\circ}$C or 80\,nm Al$_2$O$_3$ grown by atomic layer
deposition. In both cases TiAu was deposited as metallic gate.
Magnetotransport measurements were performed at temperatures $T$
between 1.5\,K and 15\,K and in magnetic fields $B$ up to 10\,T.
Several devices from the same wafer have been studied.


\begin{figure}
\includegraphics[width=\columnwidth, keepaspectratio]{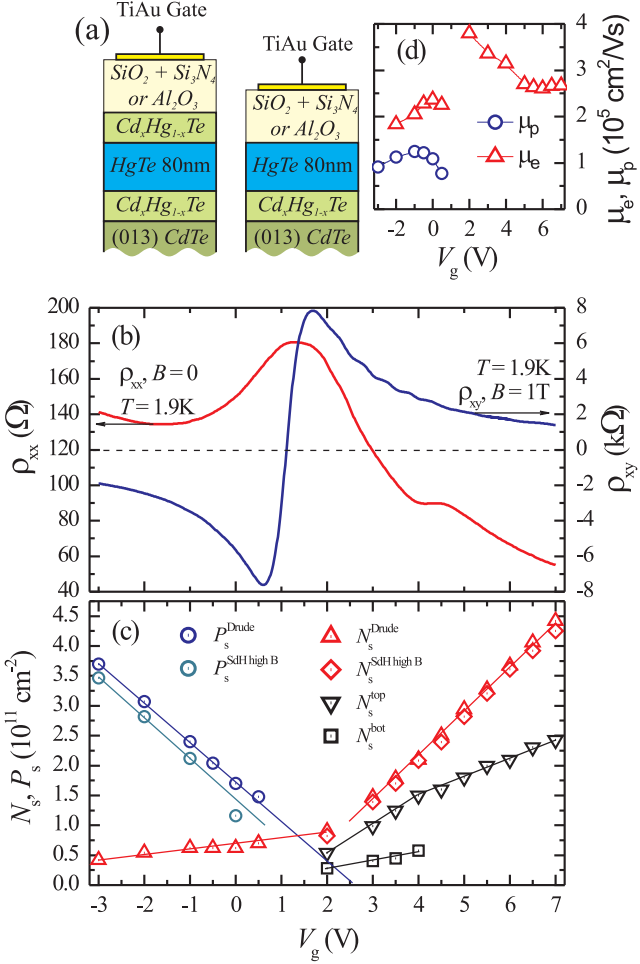}
\caption{\label{fig1} (a) Cross-section of our heterostructures.
(b) Left vertical axis: $V_g$ dependence of $\rho_{xx}$ at $T =
1.9$\,K and for $B=0$.  Right vertical axis: Hall resistance
$\rho_{xy}(V_g)|_{B=1T}$ for the second sample. (c) Electron
$N_s(V_g)$ and hole $P_s(V_g)$ densities at different $V_g$
extracted from the Drude model and high-field SdH oscillations.
The electron density of the top  layer, $N_s^{\rm top}$ is
extracted from low-field SdH oscillations (see text).
(d) Electron and hole mobilities, averaged over all participating
charge carriers, at different gate voltages. Between $V_g =
0.5$\,V and $2$\,V the fits to the two-carrier Drude model were
not reliable enough to extract electron and hole mobilities (see
supplement).}
\end{figure}


Fig.~1(b) shows the typical resistivity $\rho_{xx}$ at $B =
0$ and Hall resistance $\rho_{xy}$ at $B = 1$\,T as a
function of gate voltage $V_g$ at $T = 1.9$\,K for a HgTe film
with Si$_3$N$_4$ insulator. The $\rho_{xx}$ trace
exhibits a maximum near $V_g = 1$\,V and is asymmetric with
respect to the gate voltage: the resistance on the left hand side
of the maximum is significantly higher than on the right side.
While $\rho_{xx}$ displays a maximum in Fig.~1(b), $\rho_{xy}$,
taken at 1\,T, changes sign at the same $V_g \sim 1$\,V. This
suggests that the Fermi level can be tuned, as a function of
$V_g$, from the conduction band to the valence band.

For $V_g < 1$\,V the Fermi level is in the valence band, where
according to band structure calculations (ref. \cite{Brune11},
supplement), holes and Dirac electrons coexist. The coexistence of
two types of charge carriers is experimentally supported by a
large positive magnetoresistance $\rho_{xx}(B)$ and by a
non-linear $\rho_{xy}(B)$, typical for electron-hole systems
\cite{Blatt68, Kvon08}. To estimate mobility and density of the
coexisting electron and holes we used the Drude formalism for two
types of carriers. The total conductivity $\sigma_{xx}$ and Hall
conductivity $\sigma_{xy}$ is given by the sum of electron and
hole contributions to the conductivity.
After tensor inversion one obtains expressions for $\rho_{xx}(B)$
and $\rho_{xy}(B)$ which can be fitted to the experimental data at
each $V_g$ value. Examples are given in the supplement. A similar
analysis to extract densities and mobilities of the two carrier
species has been employed previously in semimetallic HgTe quantum
wells \cite{Kvon08}. At $V_g \geq 2$\,V the non-linear Hall effect
turns into an essentially linear dependence of $\rho_{xy}(B)$
indicating that the current is carried by only one sort of charge
carriers, i.e. surface electrons. A remaining small non-linearity
of $\rho_{xy}(B)$ is due to different densities and mobilties of
these electrons on top and bottom surface.

The resulting electron and hole 2D density and mobility are
presented in Fig.~1(c) (labeled as "Drude") and (d) \footnote{Note
that the experiment delivers averages over both surfaces and, in
certain voltage regions, also over part of the bulk. This can be
seen from the band scheme (Fig. S5, supplement) which shows the
position of $E_F$ at different $V_g$}. Both, electron and hole
densities change with $V_g$ by a factor of nearly 10, thus
indicating that only a small concentration of bulk impurities
contributes to the conductivity. Fig.~1(d) shows that electrons
and holes exhibit very high mobilities exceeding
$10^5$\,cm$^2$/V$\cdot$s. The maximum electron mobility of $\mu =
4\times10^5$\,cm$^2$/V$\cdot$s is an order of magnitude higher
than in strained HgTe films discussed previously
\cite{Brune11,Olshanetsky11} and almost a thousand times higher
than in 3D Bi TI samples. Linear extrapolation of the $P_s^{\rm
"Drude"}(V_g)$ data in Fig.~1(c) gives an intercept at $V_g
\approx 2.5$\,V, suggesting that the valence band filling starts
around here.  At about the same value of $V_g$ the slope of
$N_s^{\rm "Drude"}(V_g)$ in Fig.~1(c) and the temperature
dependence of $\rho_{xx}(V_g)$, shown in Fig.~2(a), change
noticeably: For $V_g > 2$\,V $\rho_{xx}$ does hardly change with
$T$ but changes dramatically for $V_g < 2$\,V where $\rho_{xx}$
varies by a factor 2 between $T=1.9$\,K and 15\,K. This behavior
is ascribed to strong Landau scattering \cite{Gantmakher87} of
coexisting electron and hole states, similar to the one observed
in \cite{Olshanetsky09}.

Another feature in $\rho_{xx}(V_g)$ emerges at $V_g = 4$\,V
(Fig.~1(b)), accompanied by a  change of the slope of $N_s^{\rm
top}(V_g)$ (Fig.~1(c), extracted from low-field Shubnikov-de Haas
(SdH) oscillations (see below). We suggest that this features mark
the gate voltage at which the Fermi level starts to enter the
conduction band. Therefore, the data presented in Fig.~1 and
Fig.~2a imply that the gap opens between $2$ and $2.5$\,V (top of
valence band) and closes around $4$\,V (bottom of conduction
band). Then only Dirac states localized at the two surfaces of the
strained 80\,nm HgTe film contribute to transport. Outside this
$V_g$ region Dirac electrons and bulk electrons (holes) conduct in
parallel. A sketch of the corresponding density of states versus
energy is shown in the inset of Fig.~2(a). Using the electron
densities extracted at $E_v$ and $E_c$ and the calculated k-linear
dispersion of Dirac electrons \cite{Brune11} we estimate a gap
size of $\approx 15$\,meV. This value is very close to the one
calculated in \cite{Brune11} for a strained HgTe film. The same
value we found for the Cd$_{0.7}$Hg$_{0.3}$Te capped HgTe film,
indicating that the surface states are not affected by the precise
nature of the interface.

\begin{figure}[b]
\includegraphics[width=\columnwidth,keepaspectratio]{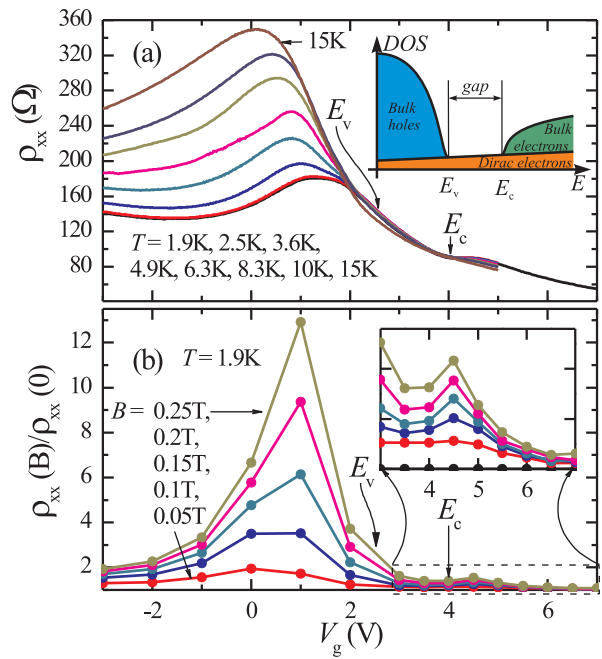}
\caption{\label{fig2} (a)  $\rho_{xx}(V_g)$ for different temperatures. The inset sketches the density of states in our system.  (b)
 $\rho_{xx}(B)/\rho_{xx}(B=0)$ vs. $V_g$ for different  magnetic field values. "$E_v$"
and "$E_c$" mark the onset of valence and conduction band. }
\end{figure}


To check the validity of our picture further we resort to
magnetoresistance measurements in a perpendicular magnetic field,
displayed in Fig.~2(b). The  magnetoresistance (MR) within
classical Drude theory for two groups of carriers (labeled by
index 1 and 2) is, for small fields, proportional to $B^2$. Its
magnitude, normalized to $\rho_{xx}(B = 0)$ is proportional to
$\frac{\sigma_1 \sigma_2}{(\sigma_1 + \sigma_2)^2}(\mu_1 \mp
\mu_2)^2$, where $\sigma_i$ and $\mu_i$ are conductivity and
mobility of the respective carrier species $i$ at $B=0$. The sign
between $\mu_i$ depends on the carriers' polarity: in case the
carriers have the same polarity (e.g. Dirac electrons in the
conduction band) the "-" sign applies while in the case of
different polarity (electrons and holes coexist) the mobilities
add, thus resulting in a strong MR. Hence the maximum of the MR
magnitude is expected to occur in the semimetal state with the
Fermi level residing in the valence band near the charge
neutrality point, where $N_s \approx P_s$ and  $\sigma_1$ and
$\sigma_2$ become comparable. A significant change of the MR
magnitude is expected when the Fermi energy moves from the valence
band into the gap where only Dirac electrons reside. For this
single carrier type the MR is small. Corresponding normalized
$\Delta \rho_{xx}(V_g)$ and $\Delta \rho_{xx}(B)$ data, displayed
in Fig.~2(b), are in accord with this expectation. For $V_g <
2$\,V a large parabolic MR is observed with a pronounced MR
maximum at $V_g = 1$\,V. In contrast, for $V_g > 2$\,V, i.e. where
Dirac electrons prevail, the MR drops by a
factor of up to 10. The MR is expected to rise again when the Fermi
level moves from the gap into the conduction band, as two (or more) groups of carriers with different
mobility are involved, i.e. Dirac and bulk electrons. Bulk
electrons near the bottom of the conductance band are expected to have a
much lower mobility than Dirac electrons. A corresponding increase
is indeed visible in Fig.~2(b) (marked by "$E_c$").

Important extra information can be obtained from experiments in
quantizing magnetic fields. SdH oscillations
and quantized Hall steps can be seen in Fig.~3(a) as a function of
$V_g$ for $B=4$\,T. The maximum in $\rho_{xx}$ at $V_g \sim 1$\,V
corresponds to the charge neutrality point at which the Hall
resistance (voltage) changes sign. On the left hand side, where
low mobility holes dominate, no quantum Hall steps appear. In
contrast, on the right hand side where Dirac fermions and
conduction band electrons prevail, quantized Hall steps develop.

Surprisingly, the Hall steps extend into a $V_g$ region where
conduction band and Dirac electrons coexist. The electron density,
extracted from high-field SdH oscillations, plotted in Fig.~1(c),
is identical to the one obtained from classical Drude fits. This
means that the filling factors at high $B$-fields are determined
by the total electron concentration $N^{\rm tot}_s$, i.e. bulk and
surface electrons. Similarly, at large negative $V_g$ and high
$B$, the hole density extracted from SdH oscillations and from
Drude theory are nearly equal, while at smaller bias the SdH data
deliver smaller hole densities. This suggests that the filling
factor in the valence band is given by the difference of bulk hole
and surface electron density. This is similar to the situation
observed in GaSb/InAs heterojunctions where electrons and holes
coexist \cite{Mendez85}.  A central observation is thus that bulk
(which in our case is a 80 nm wide HgTe quantum well) and surface
charge carriers determine jointly the high-$B$ LL filling factors.
The $\rho_{xx}(B)$ minima do not vanish indicating parallel
conduction probably due to the sides of the HgTe layer, which are
oriented parallel to the applied magnetic field \cite{Brune11}.
The quantized Hall steps on the electron side display, as in
Br\"{u}ne's work  \cite{Brune11}, even and odd integer plateau
values, thus indicating different carrier densities for top and
bottom surface.

\begin{figure}[b]
\includegraphics[width=\columnwidth,keepaspectratio]{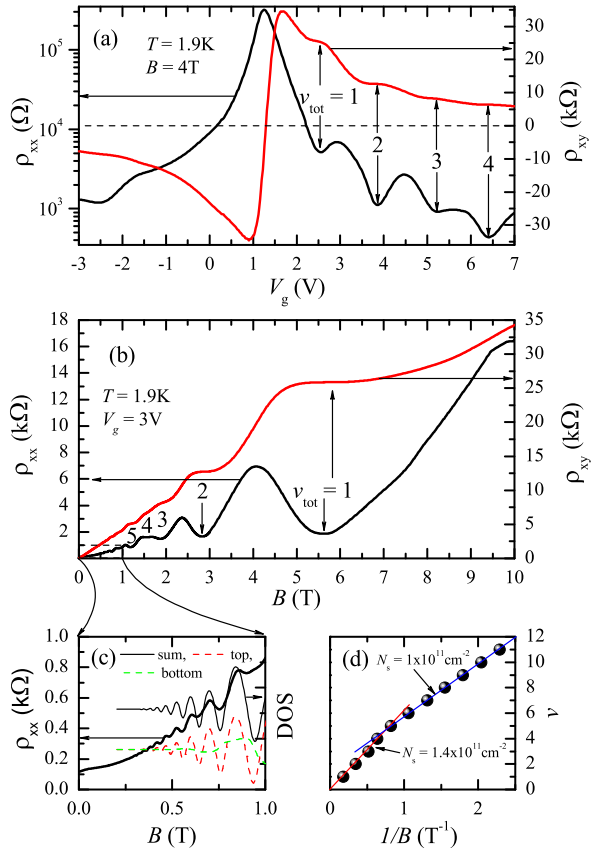}
\caption{\label{fig3} (a) $V_g$ dependence of $\rho_{xx}$
and  $\rho_{xy}$ at 4\,T. Numbered arrows mark the total filling
factors $\nu_{\rm tot}$. (b) Magnetoresistance $\rho_{xx}$ and
Hall resistance $\rho_{xy}$ for  $V_g=3$\,V, i.e. in the TI
regime. (c) Left axis: Magnification of $\rho_{xx}(B)$ at low
fields. Right axis: calculated DOS for $N_s^{\rm top} =
1.015\times10^{11}$\,cm$^{-2}$ and $N_s^{\rm bot} =
0.4\times10^{11}$\,cm$^{-2}$. (d) $\rho_{xx}(B)$ minima positions
on a  $1/B$ scale with the corresponding densities.}
\end{figure}

This is not surprising as it is a consequence of screening, i.e.,
a part of the electric field gets screened by the top layer of
Dirac electrons. Applying an electric field to the top gate hence
results in different filling rates $d N_s^{\rm top (bot)}/dV_g$.
These rates can be easily estimated if $E_F$ is in the gap, i.e.
between $\sim 2$\,V and $4$\,V. Then the change of top and bottom
electron density
is given by \cite{Luryi88} $\Delta N_s^{\rm top}/\Delta N_s^{\rm
bot} = 1 + \frac{q^2 D d_{\rm HgTe}}{\varepsilon_{\rm HgTe}
\varepsilon_0}$, where $D$ is the density of states of Dirac
electrons on the top surface, and $d_{\rm HgTe}$ and
$\varepsilon_{\rm HgTe}$ are thickness and dielectric constant of
the HgTe layer, respectively. Inserting typical values we obtain
$\Delta N_s^{\rm top}/\Delta N_s^{\rm bot} = 3-5$. Below we show
that the experimentally observed difference in top and bottom
layer filling rate is close to this expected value. Besides
$N_s^{\rm tot}$ we plot in Fig.~1(c)  also the electron density of
the top surface as a function of $V_g$. The corresponding data are
obtained from the following consideration: Assuming that the
carrier densities are equal at the flat band condition, i.e. at
$V_g = 0$\,V, the electron density on the top surface becomes
significantly higher for $V_g > 2$\,V. The higher carrier density
of the top surface is expected to be connected with a higher
electron mobility \footnote{Comparison of CdHgTe capped and
uncapped heterostructures (see supplement for details) proves that
they have the same average mobility for $E_F$ in the gap.  We thus
conclude that for the same carrier density mobilities on top and
bottom surface are approximately the same}. This offers an
opportunity to separate the electron density of top and bottom
layer  experimentally. As the higher mobility of the top layer is
connected to a smaller Landau level broadening, Shubnikov de Haas
oscillations commence at lower magnetic fields and dominate the
low field magnetoresistance oscillations in Fig.~3(c). This is
reflected in different periods of SdH oscillations in low and high
$B$-fields, displayed in Fig.~3(d). The $1/B$ positions of the SdH
minima vs. filling factors can be fitted by two straight lines
corresponding to the carrier density $N_s^{\rm top} =
1\times10^{11}$\,cm$^{-2}$ of the top layer and
the total carrier density $N_s^{\rm tot} 
=1.4\times10^{11}$\,cm$^{-2}$. $N_s^{\rm top}$, extracted from
low-field SdH-oscillations for different $V_g$ is shown in
Fig.~1(c). In both cases we assumed spin-resolved LLs. The reduced
slope of $N_s^{\rm top}(V_g)$ for $V_g > 4$\,V is a clear
signature that $E_F$ moves in the conduction band.  Since
$dN_s^{\rm tot}/dV_g$ is constant, $dN_s^{\rm top}/dV_g$ decreases
when the bulk electron density $N_s^{\rm bulk}$ starts to appear,
i.e., when $dN_s^{\rm bulk}/dV_g > 0$. For $B > 1$\,T Landau
quantization gets resolved in the lower mobility bottom layer, too
and, due to electron redistribution, the two surfaces (and for
$V_g > 4$\,V also the bulk electrons) act like a single 2DEG with
density $N_s^{\rm tot}$. The redistribution of electrons in the TI
state is possible via contacts, ungated conducting regions and via
side facets of the HgTe film.

With $N^{\rm tot}_s$, acquired from high-field SdH oscillations,
and $N_s^{\rm top}$ we can, for $E_F$ between $2$\,V$<V_g<4$\,V,
i.e. in the TI regime, calculate the carrier density of the bottom
layer, $N_s^{\rm bot}$. Corresponding data are also shown in
Fig.~1(c). The slope of $N_s^{\rm top}(V_g)$ is by a factor of
$\sim 3$ higher  than the one of $N_s^{\rm bot}(V_g)$. This is in
line with the effect of screening, discussed above.

In summary we have shown that an analysis of magnetotransport data
in strained high-mobility HgTe layers brings out the different
carrier types contributing to transport at different Fermi level
positions. A strong magnetoresistance, observed in the valence
band where electrons and holes coexist, is absent in the gap where
only surface electron contribute, and rises again when the Fermi
energy enters the conduction band. An analysis of high- and
low-field quantum oscillations highlights the interplay of the
different carrier types and allows to probe the carrier density of
top and bottom layer in the TI regime separately.

Authors acknowledge the assistance of C. Linz in Al$_2$O$_3$
insulator and gate fabrication. This work was supported by the
German Science Foundation (DFG) via Research Unit 1483 and SPP
1666, by the International Bureau of the German Federal Ministry
of Education and Research (RUS 09/29), the Russian Foundation for
Basic Research, and the Russian Academy of Sciences (Physics and
Technology of Nanostructures Program).

\nocite{}

\bibliography{3dTI}%

\newpage

\onecolumngrid

\begin{center}
\Large{\textbf{Supplemental material for ``Transport properties of
a 3D topological insulator based on a strained high mobility HgTe
film"}}

\large{D.\,A.~Kozlov, Z.\,D.~Kvon, D.~Weiss, E.\,B.~Olshanetsky,
N.\,N.~Mikhailov, S.\,A.~Dvoretsky}
\end{center}

\addtocounter{figure}{-3}

\renewcommand{\thefigure}{S\arabic{figure}}

\section{Samples details}

\subsection{Heterostructure and Hall bar}

We fabricated and investigated several devices based on two types
of heterostructures. Both types were grown grown by molecular beam
epitaxy on a (013)-oriented CdTe substrate and have very similar
structure: the essential part is a strained 80\,nm thick HgTe film
(Fig.~S1(a)). The two  heterostructures types differ  in the top
layer sequence: in the first structure (named "uncapped") the HgTe
is uncapped on top while the second one ("capped") is covered with
a 20\,nm Cd$_{0.7}$Hg$_{0.3}$Te cap layer. Since HgTe films grown
on CdTe suffer from dislocations due to  lattice mismatch, thus
causing low mobility ($<4\times10^4$\,cm$^2$/V$\cdot$s) and high
impurity concentration ($>10^{17}$\,cm$^{-3}$) in the bulk
\cite{Brune11, Olshanetsky11}, our 80\,nm thick HgTe films were
separated from the CdTe substrate by a 20\,nm thin
Cd$_{0.7}$Hg$_{0.3}$Te buffer layer (Fig.~S1(a)). The buffer layer
increases the electron mobility by an order of magnitude (up to
$4\times10^5$\,cm$^{2}$/V$\cdot$s) and reduces the bulk impurity
concentration to a value about 10$^{16}$\,cm$^{-3}$, estimated
from the carrier density in ungated samples (see below).

For transport measurements the films were patterned into Hall
bars. The central part of several Hall bars is covered with a top
gate. A top view of the devices is sketched in Fig.~S1(b). Each
Hall bar consists of three 50\,$\mu$m wide segments of different
lengths (100, 250, and 100\,$\mu$m) with eight voltage probes. The
ohmic contacts to the active layer were formed by alloying indium.
In our devices we used two types of dielectric layers for gating:
100\,nm SiO$_2$ and 200\,nm of Si$_3$N$_4$ grown by plasma
chemical vapor deposition of SiH$_4$ + N$_2$O at 100$^{\circ}$C or
80\,nm Al$_2$O$_3$ grown by atomic layer deposition. In both cases
TiAu was deposited as metallic gate.

\begin{figure}[h]
\includegraphics[width=0.6\columnwidth,keepaspectratio]{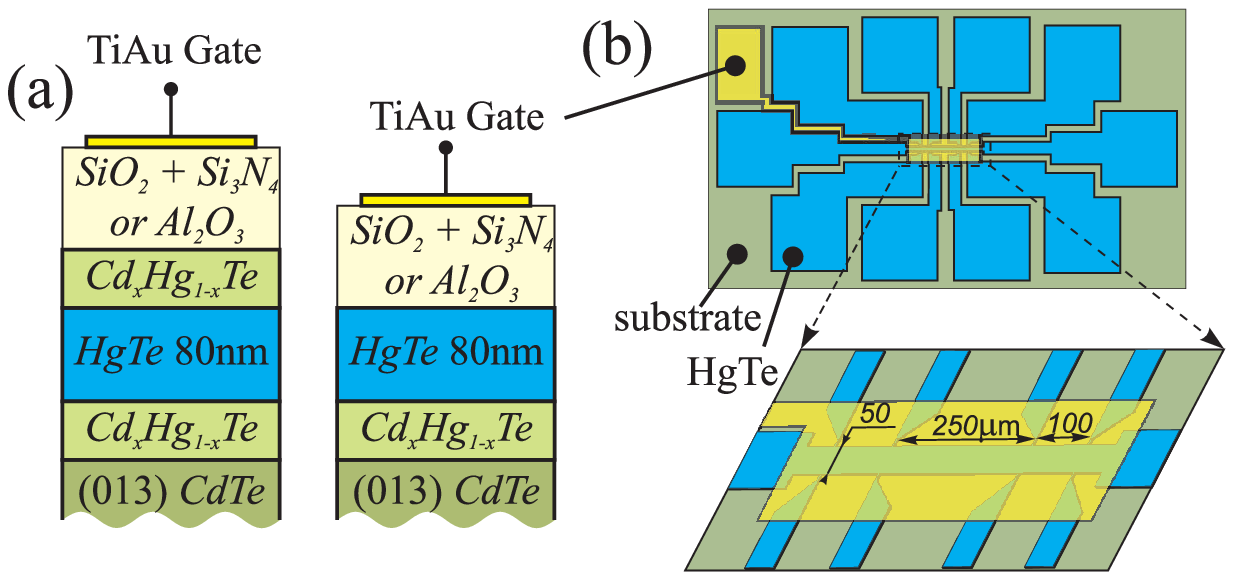}
\caption{\label{fig1} (a) Cross-section of devices with top gate.
We fabricated and investigated also ungated devices made from both
kinds of structures. (b) Schematic top view of the Hall bars
studied. }
\end{figure}

\subsection{Ungated sample properties}

Fig.~S2(a) displays $\rho_{xx}(B)$ and $\rho_{xy}(B)$ traces of
one of the ungated samples. These traces are typical for a system
with coexisting electrons and  holes. Fitting these traces by
using the classical two-component
 Drude model (see next paragraph)
allows to determine the values of 2D electron and hole densities
and mobilities. Electron and hole densities of the $80$\,nm thick
HgTe layer are expressed as 2D carrier densities. Values of
carrier densities and mobilities extracted from such fits are
presented in the boxes underneath the figures in  Fig.~S2. In
comparison with previous results \cite{Brune11, Olshanetsky11} our
samples exhibit much smaller carrier densities of typically
($(1-2)\times10^{11}$\,cm$^{-2}$) and higher mobilities ($>
10^5$\,cm$^{2}$/V$\cdot$s). Both, density decrease and mobility
increase indicate that the amount of charged impurities has been
significantly reduced by introducing of the CdHgTe buffer layer.
The amount of charged impurities in the structure can be estimated
from charge neutrality, i.e. $N_{\rm imp} \approx |N_s - P_s| =
1.6\times 10^{11}$\,cm$^{-2}$. Taking the thickness of the HgTe
layer into account this results in  a
 volume impurity density of $N_{\rm imp}^{3D}\approx
2\times10^{16}$\,cm$^{-3}$.

\begin{figure}[h]
\includegraphics[width=0.85\columnwidth,keepaspectratio]{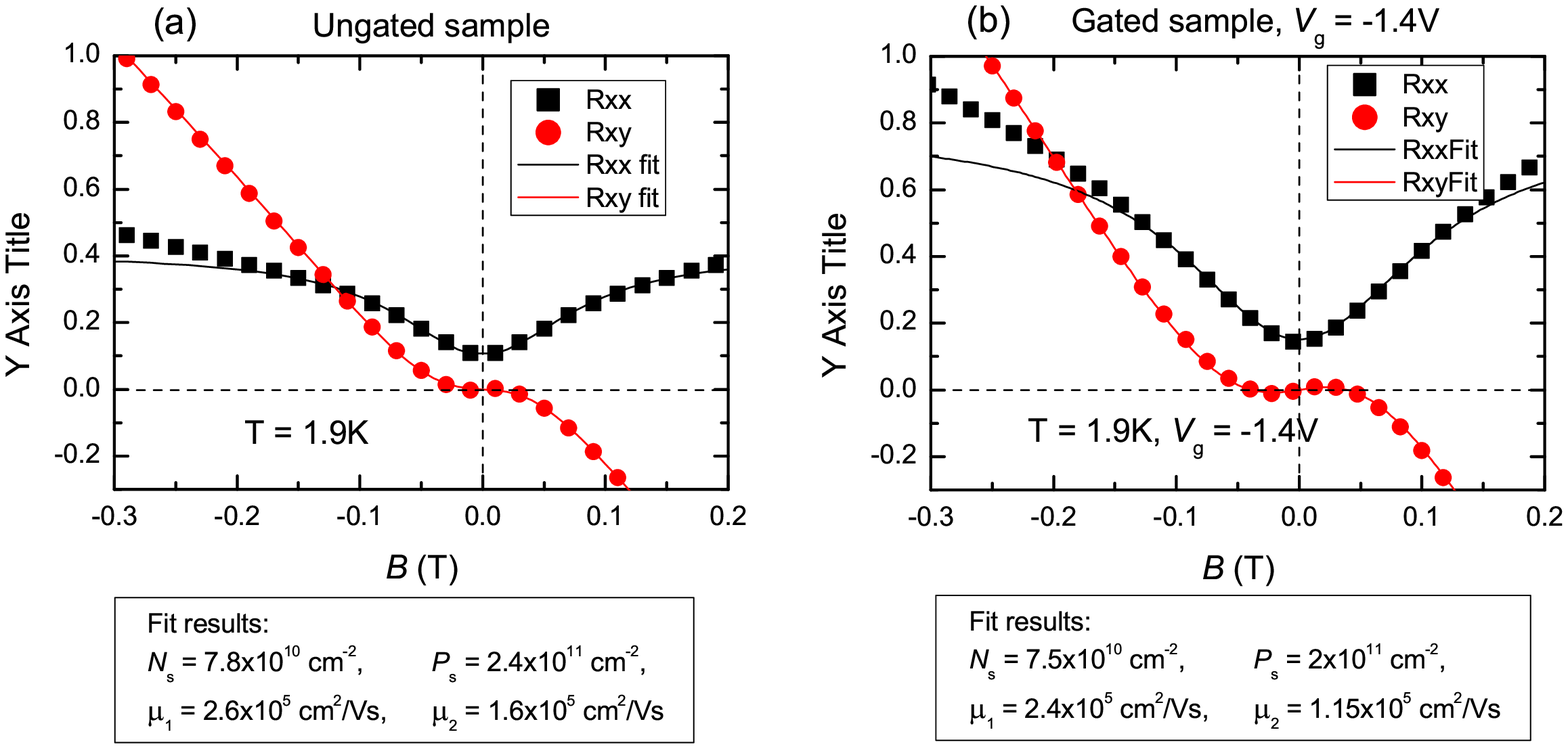}
\caption{\label{fig2} $\rho_{xx}(B)$ (black dots) and Hall
resistance $\rho_{xy}(B)$ (red dots) of ungated (a) and gated (b)
samples. Solid lines are fits to the Drude model for a
two-component 2D electron-hole gas. Extracted electron and hole
densities and mobilities are shown in the boxes underneath. }
\end{figure}

\subsection{Gated and ungated samples}

Whether the Dirac surface states are affected by gate fabrication
we studied  with some care. In Fig.~S2 we compare $\rho_{xx}(B)$
and $\rho_{xy}(B)$ of the ungated sample (Fig.~S2(a)) and of the
gated one (Fig.~S2(b)) for an applied gate voltage $V_g =
-1.4$\,V. The most important feature here is the overall
similarity of the corresponding traces in  Fig.~S2(a) and (b): by
applying an appropriate gate voltage we can bring the gated system
into the same state as the ungated one. Direct comparison of
electron and hole densities and mobilities, extracted from Drude
fits, shows that they are  within 20\% the same. This is the same
level of difference found between samples fabricated the same way.
This indicates that gate fabrication does not change the band
diagram of the structure qualitatively but introduces charged
impurities and/or defects in the system with a density of order of
$(1-2)\times10^{11}$\,cm$^{-2}$, varying from sample to sample.

\subsection{CdHgTe-capped and open HgTe films}

\begin{figure}[h]
\includegraphics[width=0.7\columnwidth,keepaspectratio]{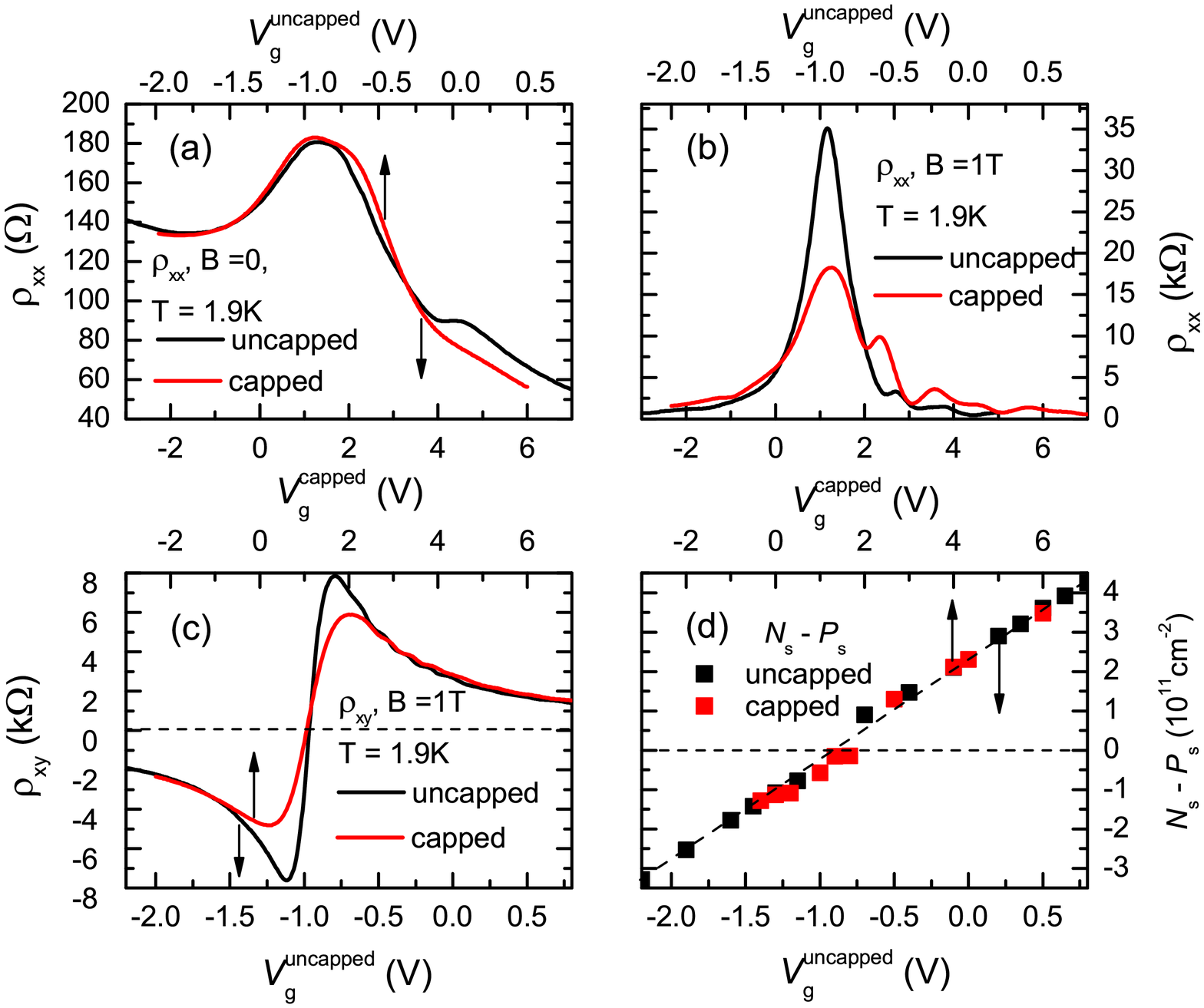}
\caption{\label{fig3} Comparison of the gated sample with uncapped
HgTe film and with SiO$_2$+Si$_3$N$_4$ insulator ("uncapped",
black lines) and the sample with the HgTe film capped by a CdHgTe
layer and with Al$_2$O$_3$ insulator ("capped", red lines). (a)
Resistivity $\rho_{xx}$ vs gate voltage at zero magnetic field and
$T = 1.9$\,K. (b) $\rho_{xx}(V_g)$ at $B = 1$\,T. (c). Hall
resistance $\rho_{xy}(V_g)$ at $B = 1$\,T. (d) Charge $N_s(V_g) -
P_s(V_g)$ in the system  determined by fitting of $\rho_{xx}(B)$
and $\rho_{xy}(B)$ at fixed gate voltages by the two-component
Drude model.}
\end{figure}

Besides investigating the influence of the gate fabrication on the
sample's charge state we also checked how capping of the HgTe film
with a CdHgTe film  affects the transport properties (see
Fig.~S1(a)). In addition we compared different kinds of dielectric
layers under the gate. The following results show that the
properties of the HgTe film are only weakly dependent on the
presence of a cap or different dielectric layers. A comparison
between two samples is presented in the Fig.~S3. The first one
(labeled "uncapped" in Fig.~S3) has an uncapped HgTe film with an
insulating SiO$_2$ layer on top and the second one ("capped" in
the Fig.~3) is a HgTe film with CdHgTe cap layer and a Al$_2$O$_3$
insulator under the gate. Since dielectric constant and insulator
thickness are different for these samples  the change of carrier
density $N_s-P_s$ with gate voltage $V_g$ differs correspondingly:
$d(N_s-P_s)/dV_g ^{\rm Si_3N_4} \approx 0.75 \times
10^{11}$\,cm$^{-2}$/V for the Si$_3$N$_4$-covered sample and
$d(N_s-P_s)/dV_g^{\rm Al_2O_3} \approx 2.5 \times
10^{11}$\,cm$^{-2}$/V for the Al$_2$O$_3$-covered one. In order to
compare the samples one has to rescale the gate voltage axis with
the ratio of the respective specific capacitances. In doing so,
the experimental traces in  Fig~S3(a), (c) and (d) of the two
samples follow each other closely.  Only some features in
$\rho_{xx}$ and $\rho_{xy}$ in the presence of a  magnetic field
and near the charge neutrality point in Fig.~S3(b) and (c)
 differ slightly. In spite of these small differences the general
properties of the samples are still the same: all the observed
features in transport
 of the uncapped sample are more or less visible in
the measurements of the capped sample and even the estimated gap
value between conduction and valence bands is the same -- $\Delta
\approx 15$\,meV. The reproducibility between the two samples made
from different wafers is amazing and points to some kind of
rigidity of the band structure and Dirac surface states with
respect to external disturbances.

\section{Fitting of magnetoresistance and Hall resistance within the two-component classical Drude model}

\begin{figure} [h]
\includegraphics[width=1\columnwidth,keepaspectratio]{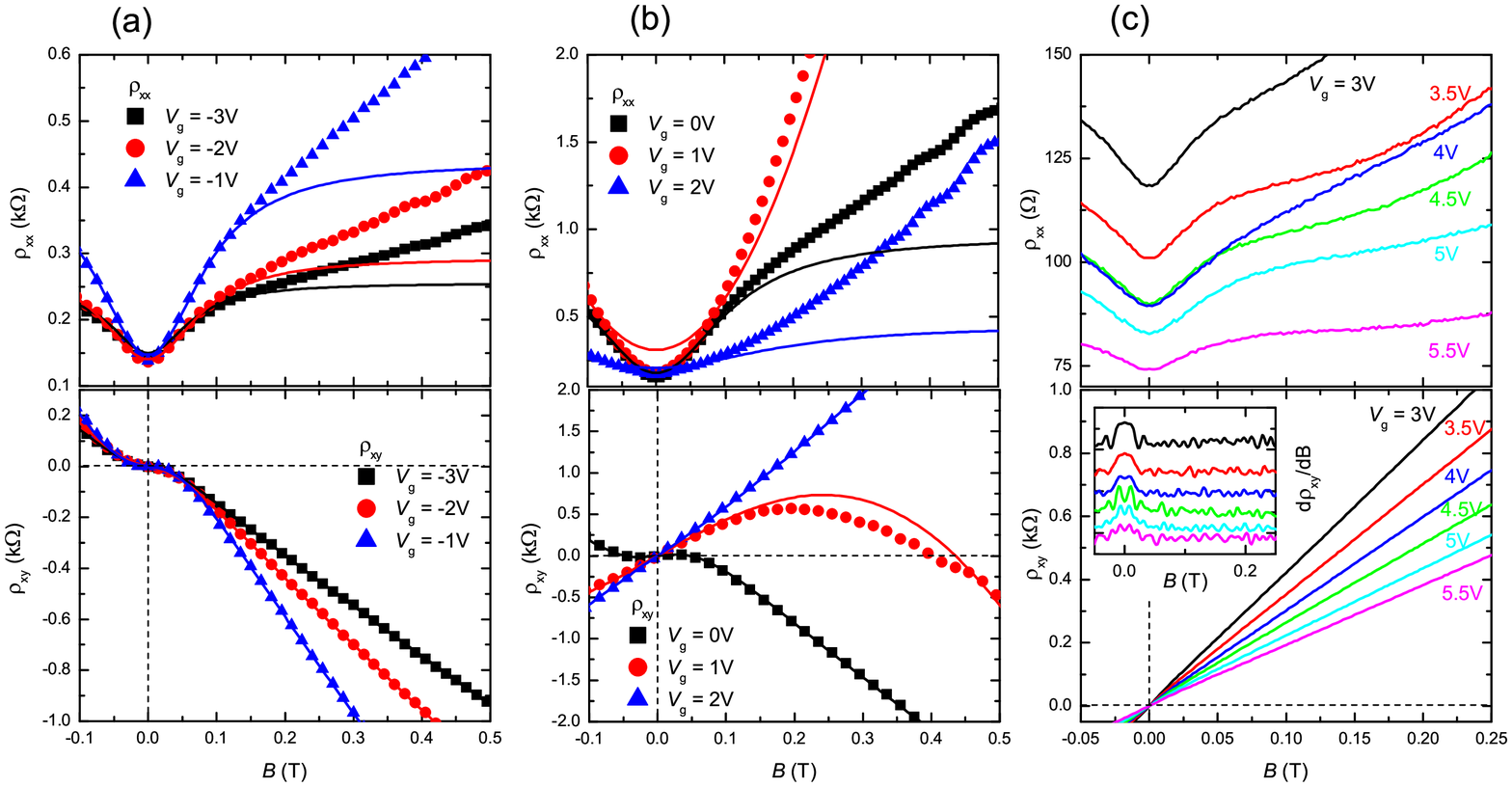}
\caption{Resistivity $\rho_{xx}(B)$ (top) and Hall resistance
$\rho_{xy}(B)$ (bottom)  for different gate voltages. Fits to the
two-component Drude model are drawn as solid lines. In the
semimetal state (a) ($V_g =  -3...0$\,V) it is possible to fit
$\rho_{xy}$ over the $B$-range and $\rho_{xx}$ around $B=0$. (b)
Gate voltage region corresponding to the transition from the
valence band (semimetal state) to the topological insulator state.
Between $V_g = 0$\,V and $V_g = 2$\,V the fits work less well,
most probably to to the proximity to the charge neutrality point,
at which small inhomogeneities are more detrimental. (c)
Magnetoresistance and Hall resistance in the topological insulator
region  ($V_g = 2 ... 4$\,V) and at the transition to the bulk
electron metal ($V_g > 4$\,V). For all gate voltages
$\rho_{xy}(B)$ shows essentially a linear dependence on $B$. This
indicates that the current is carried by only one sort of charge
carriers. A more detailed analysis of $\rho_{xy}(B)$ shows a
slight nonlinearity (see derivative $d\rho_{xy}(B)/dB$ in the
insert) near $B =0$ indicating  the presence of two (two surfaces
with Dirac electrons in the TI state) or more (Dirac electrons on
top and bottom surface and bulk electrons in the conduction band)
groups of carriers with the same charge but with different
mobilities. However the magnitude of the nonlinearity is so small
making it impossible to distinguish these different groups
reliably within the Drude model. For  $V_g \geq 2$\,V we
determined only the overall electron density and their average
mobility without differentiation between several groups.}
\end{figure}

We used the classical Drude model  to extract electron and hole
densities and mobilities by fitting $\rho_{xx}(B)$ and
$\rho_{xy}(B)$  at small magnetic field and for a fixed gate
voltage (see Fig.~S4). A detailed description of the  model can be
found in \cite{Blatt68} and many other textbooks. Here we present
only  its application to our system.  The model is based on the
assumption that, in the system studied, $n$ groups of
noninteracting carriers exist. Each group $i=1...n$ has charge
$q_i=\pm q$ (electrons or holes), density $n_{2D}^i$ and mobility
$\mu_i$. The conductivity of the system is given as the sum over
the conductivities of each group: $\sigma_{xx} = \sum_i^n
\sigma_{xx}^i$ and $\sigma_{xy} = \sum_i^n \sigma_{xy}^i$. After
tensor inversion one  obtains the corresponding expressions for
$\rho_{xx}(B)$ and $\rho_{xy}(B)$.

\begin{figure}[h]
\includegraphics[width=0.85\columnwidth,keepaspectratio]{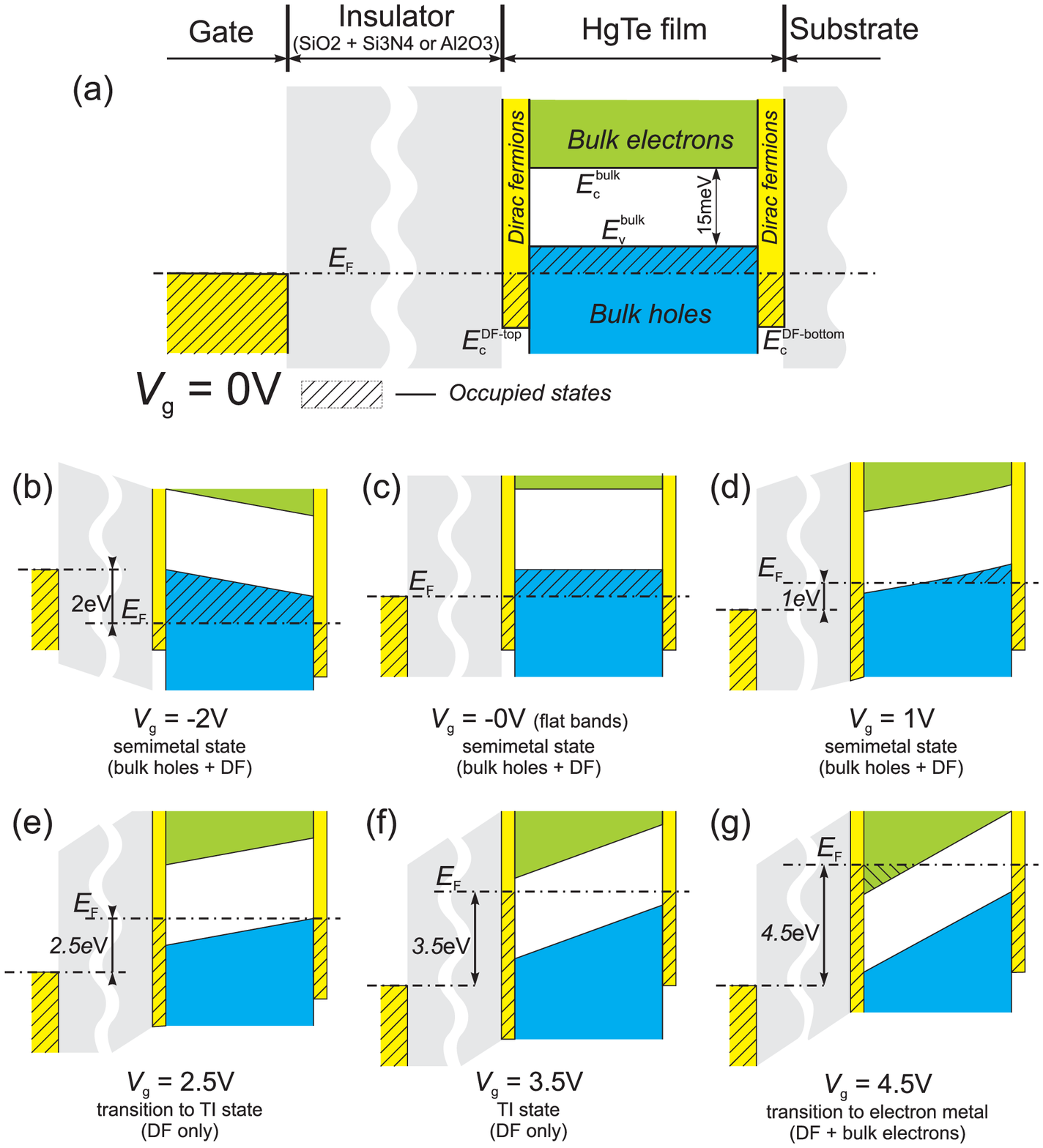}
\caption{\label{fig3} Schematic band diagrams of our system for
different gate voltages. The Dirac point is located in the valence
band indicated by the onset of the yellow bars which illustrate
the Dirac electrons on top and bottom surface. In the band
structure shown here, flat band occurs at $V_g = 0$\,V. Applying a
bias voltage results in band bending. When $E_F$ is in the gap in
(e) and (f) the screening on the electron density of the back
surface can be estimated (see main text). The band diagram under
different bias condition show that the electron (hole) densities
and mobilities are always averages over top and bottom surfaces
and, depending on bias, also involving bulk electrons (see (g))
and bulk holes (see (b)). }
\end{figure}

In our system we can distinguish four main groups of carriers:
bulk holes in the valence band, bulk electrons in the conduction
band and two groups of Dirac fermions (DF) which exist at the two
surfaces of the HgTe film. This picture  is based on band
structure calculations of Br\"{u}ne et al. (\cite{Brune11},
supplementary). A simplified band scheme which describes the
position of the Fermi energy at different gate voltages $V_g$ is
presented in Fig.~S5. Though the Drude model can, in principle, be
used for any number $n$ of carrier groups,  a reasonable fit is
only possible for $n=1$ (trivial case) and $n=2$. Already for
three groups of carriers the expressions become complicated and
contain 6 fitting parameters. Then fitting become ambiguous. Due
to the gap in the bulk spectrum we have in our case no more than
three groups of carriers at $E_F$ (see Fig.~S5). To reduce the
number of groups further we treat the Dirac electrons of top and
bottom surface as one group. This is justified as they have the
same charge and since the difference in mobility and carrier
density causes only weak corrections to $\rho_{xx}(B)$ and
$\rho_{xy}(B)$. When extracting carrier density and mobility of
surface electrons we always obtain an average of top and bottom
surface. Within this approximation the number $n$ of Drude groups
is reduced to 2 (bulk holes and Dirac electrons or bulk electrons
and Dirac electrons when $E_F$ is located in the valence or
conduction band, respectively). With $E_F$ in the bulk gap $n$ is
even reduced to one as then only Dirac electrons exist.

In case of two of groups of carriers (with index 1 and 2) the
system displays a classical parabolic magnetoresistance (MR) with
magnitude $d^2 \rho_{xx}(B) /dB^2|_{B=0} = \frac{\sigma_1
\sigma_2}{\sigma_1 + \sigma_2}(\mu_1 \mp \mu_2)^2$, where
$\sigma_i$ and $\mu_i$ are the corresponding  conductivities of
each group $i$ at zero magnetic field. The sign between the
$\mu_i$'s depends on the carriers' polarity: in case the carriers
have the same polarity (e.g., two groups of electrons)  the "-"
sign applies while in the case of different polarity (electrons
and holes coexist when Fermi level resides in the valence band)
the mobilities add, thus resulting in a strong MR. Hence two
groups of Dirac electrons with similar values of their mobility
will cause only a small MR effect. This is actually the reason why
we can treat top and bottom surface electrons as one species. In
contrast, we expect a maximum magnetoresistance when the
 Fermi level is in the valence band near the
the charge neutrality point where $N_s \approx P_s$ and $\sigma_1
\approx \sigma_2$ holds. Note that the charge neutrality point is
not equal to the Dirac point as the latter is located in the
valence band. A significant change of the MR magnitude is expected
when the Fermi level moves from the valence band into the gap
where only Dirac electrons reside (topological insulator state)
and the magnitude of MR drops by up to a factor of 10. The data in
Fig.~S4(b) show this drastic chance in MR depending on $V_g$, i.e.
$E_F$. The drastic change of $\rho_{xy}(B)$ between $1$\,V and
$2$\,V in Fig.~S4(b) also heralds this change in the contributing
charge carriers. At $2$\,V $\rho_{xy}(B)$ becomes nearly linear
indicating that for $V_g > 2$\,V only the electrons of the surface
states contribute.

The next change of the MR behavior is expected when the Fermi
level enters the conduction band. (Fig.~S4(c)).  There, the
presence of low-mobility bulk electrons results in a rise of the
MR magnitude (see, e.g., $\rho_{xx}(0.1 \rm T) - \rho_{xx}(0)$).

The monotonous decrease of $d^2 \rho_{xx}(B) /dB^2$ (curvature
around $B=0$) with increasing $V_g$ is reversed in the small
voltage interval between $V_g = 4$\,V and $V_g = 4.5$\,V. Because
of the relatively small bulk electron density when $E_F$ just
enters the conduction band and their small contribution to the
overall conductivity, the resulting MR is weaker compared to the
situation in the valence band where carriers of different polarity
are involved. For further increasing $V_g$ the bulk electrons'
contribution to the conductivity increases but, on the other hand,
the mobility increases also resulting in a reduced MR magnitude
for $V_g>5$\,V, as displayed in Fig.~S4(c).

Thereby the presence of the bulk electrons, starting to appear at
$\sim 4$\,V is observed in classical transport. However, the Drude
fits are not sensitive enough to disentangle density and mobility
of surface and bulk electrons and one could determine the gate
voltage corresponding to crossing of Fermi level with the bottom
of conductive band but the sensitivity of the Drude fitting
technique is not enough to determine exact values of bulk
electrons density and mobility.


\nocite{}


\end{document}